\documentclass[letterpaper, 11pt]{article}
\usepackage{amsmath}
\usepackage{amsfonts}
\usepackage{gensymb}
\usepackage{comment} 
\usepackage{listings}
\usepackage[table]{xcolor}
\usepackage{subcaption}
\usepackage{float}
\usepackage{tabu}
\usepackage{accents}

\usepackage{cite}
\usepackage{wrapfig}

\usepackage[bookmarks=false]{hyperref}

\usepackage{fullpage} 
\usepackage{graphicx}
\usepackage{enumitem}
\usepackage{makecell}

\def\CC{{\mathbb C}}

\def\NN{{\mathbb N}}

\def\ZZ{{\mathbb Z}}

\usepackage{authblk}

\begin{document}

\title{Special Function Methods for Bursty Models of Transcription}
\author[1]{Gennady Gorin}
\author[2,*]{Lior Pachter}
\affil[1]{Division of Chemistry and Chemical Engineering, California Institute of Technology, Pasadena, CA, 91125}
\affil[2]{Division of Biology and Biological Engineering \& Department of Computing and Mathematical Sciences, California Institute of Technology, Pasadena, CA, 91125}
\affil[*]{Address correspondence to Lior Pachter (lpachter@caltech.edu)}

\maketitle

%

\section{Abstract}

We explore a Markov model used in the analysis of gene expression, involving the bursty production of pre-mRNA, its conversion to mature mRNA, and its consequent degradation. We demonstrate that the integration used to compute the solution of the stochastic system can be approximated by the evaluation of special functions. Furthermore, the form of the special function solution generalizes to a broader class of burst distributions. In light of the broader goal of biophysical parameter inference from transcriptomics data, we apply the method to simulated data, demonstrating effective control of precision and runtime. Finally, we suggest a non-Bayesian approach to reducing the computational complexity of parameter inference to linear order in state space size and number of candidate parameters.

\section{Background}

Recent improvements in transcriptomics and fluorescence microscopy methods have enabled the rapid and accurate quantification of mRNA on a transcriptome-wide scale with single-molecule precision \cite{xu_stochastic_2016, lee_covering_2018, shah_dynamics_2018, eng_transcriptome-scale_2019, erhard_scslam-seq_2019, wissink_nascent_2019}. Simultaneous advances in biophysical and statistical modeling have enabled the effective discrimination of gene expression models and the determination of physical parameters from these data. The estimation of underlying parameters relies on the ability to compute the distribution of molecules for a proposed set of parameters. The Chemical Master Equation (CME) is the standard modeling framework for low-copy single-molecule kinetics, treating such systems with Markov chains traversing state spaces of integer molecule counts \cite{phillips_physical_2013, gardiner_handbook_2004, mcquarrie_stochastic_1967}. However, solutions are available only for a relatively small set of models \cite{gardiner_handbook_2004, erdi_stochastic_2014}. Furthermore, the existence of a closed-form solution does not guarantee its computational tractability.

Currently popular approaches to solving the CME can be roughly divided into three categories: simulation, matrix, and analytical methods. Simulation methods, such as the Gillespie stochastic simulation algorithm \cite{gillespie_general_1976, gillespie_exact_1977}, are easily implemented and parallelized; the sample statistics of numerous realizations asymptotically approach the statistics of the underlying process, although the speed of approach varies. Matrix methods, such as finite state projection \cite{munsky_finite_2006} or multi-finite buffers \cite{cao_accurate_2016}, rely on matrix exponentiation or eigenvalue calculation to directly solve a truncation of the infinite-dimensional CME system; however, barring convenient symmetries, these methods require a characteristic running time of roughly $O(n^3)$, where $n$ is the state space size. Finally, analytical methods directly solve the underlying system of ordinary differential equations (ODEs), e.g. using a generating function representation \cite{gardiner_handbook_2004} or a convenient basis \cite{jahnke_solving_2006}, and can be run in $O(n)$ time. 

Due to lower computational complexity, these analytical methods are highly relevant to the determination of biophysical parameters from high-dimensional, multimodal data, such as that available by modern transcriptomics and proteomics methods. Recent findings suggest that the use of joint data can provide substantial improvements to model and parameter estimation \cite{munsky_distribution_2018}, motivating the development of more efficient solvers for the CME. Current chemistries can quantify spliced and unspliced mRNA molecules \cite{shah_dynamics_2018, la_manno_rna_2018}, as well as surface proteins \cite{stoeckius_cite-seq_2017, peterson_dynamics_2017}. The following multimodal models have analytical CME solutions, as well as drawbacks limiting their direct application to biological data. 

\begin{enumerate}
    \item Combination of Poissonian solutions \cite{jahnke_solving_2006,vastola_solving_2019}: cannot be applied to proteomics, and does not explicitly model multistate genes.
    \item Constitutive mRNA and protein production \cite{bokes_exact_2012}: exact solution, but applies poorly to eukaryotic systems due to prevalence of multistate genes.
    \item Telegraph mRNA and protein production \cite{shahrezaei_analytical_2008, veerman_time-dependent_2018}:  perturbative solution that relies on time-scale separation between mRNA and protein lifetimes, and inapplicable to a large fraction of eukaryotic genes.
    \item Multi-state gene solutions with a single product \cite{zhou_analytical_2012, ham_extrinsic_2019, ham_exactly_2020}: exact solution, but does not provide information regarding downstream gene products. Current sequencing methods cannot be easily integrated with DNA accessibility testing.
    \item Bursty mRNA production and isomerization \cite{singh_consequences_2012}: exact solution, but relies on numerical integration and uses a fairly simple burst model. 
\end{enumerate}

A recent method \cite{la_manno_rna_2018} uses joint distributions of spliced and unspliced mRNA to perform short-time extrapolation on the cell landscape, motivating a more detailed treatment using stochastic biophysics. Motivated by that work, we propose a semi-analytical method for the evaluation of joint distributions resulting from the bursty transcription model \cite{singh_consequences_2012}, which describes a large fraction of mammalian genes \cite{bahar_halpern_bursty_2015, golan-lavi_coordinated_2017, dar_transcriptional_2012, suter_mammalian_2011, amrhein_mechanistic_2019}. Furthermore, we apply these models to parameter estimation, and discuss their applications to a set of burst size distributions that have not been previously solved.

\section{Methods}

We follow previous literature \cite{singh_consequences_2012} in implementing a Markov model for production, isomerization, and degradation of mRNA (\textbf{Figure 1a}). A single gene locus undergoes transcriptional bursting at a rate of $k_i$, producing $B$ nascent mRNA transcripts (pre-mRNA) per burst, with $P(B=\rho)=\alpha_\rho$. The nascent transcripts are isomerized to mature mRNA. $B$ is a random variable; if the underlying gene expression follows a two-state telegraph model with short bursts of finite size, $B$ is drawn from a geometric distribution \cite{golding_real-time_2005}. The reactions are modeled as a Poisson processes with constant rates, which enables their representation using a homogeneous continuous-time Markov chain (CTMC). $P(n,m,t)$, the law of this CTMC model, yields the probability of finding $n$ nascent and $m$ mature molecules at time $t$. 

The full set of CME ODEs is as follows:
\begin{equation}
\begin{split}
    \frac{dP(n,m,t)}{dt} =& k_i \bigg(\sum_{\rho=0}^n \alpha_\rho P(n-\rho, m, t) - P(n, m, t)\bigg) \\  
       +& \beta ((n+1)P(n+1,m-1,t) - n P(n,m,t)) \\
       +& \gamma ((m+1)P(n,m+1,t) - m P(n,m,t))
\end{split}
\end{equation}

Using the probability-generating functions (PGF) $G(x,y,t) = \sum_{n=0}^\infty \sum_{m=0}^\infty x^n y^m P(n,m,t)$ and $F(x)=\sum_{\rho=0}^\infty \alpha_\rho x^\rho$, the CME recurrence relation may be cast into the form of a single partial differential equation (PDE): 

\begin{equation}
    \frac{\partial G}{\partial t} = k(F(x)-1)G+\beta(y-x)\frac{\partial G}{\partial x} + \gamma(1-y) \frac{\partial G}{\partial y}
\end{equation}

subject to the initial condition $G(x,y,0) = \sum_{n=0}^\infty \sum_{m=0}^\infty x^n y^m P(n,m,0)$ and the normalization condition $G(1,1,t)=1$. Introducing the transformations $x=1+u$, $y=1+v$, and $G=e^\phi$ results in the following PDE:

\begin{equation}
    \frac{\partial \phi}{\partial t} = k(M(u)-1)+\beta(v-u)\frac{\partial \phi}{\partial u} + \gamma v \frac{\partial \phi}{\partial v}
\end{equation}

such that $M(u)=F(1+u)$. The solution of the PDE at time $t$ is expressed by the following integral:

\begin{equation}
    \phi(u,v,t) = k_i \int_0^t [M(U(s))-1]ds + \phi(U(t),V(t),0)
\end{equation}

Per the method of characteristics,$V(s)=ve^{-\gamma s}$, $U(s)=vfe^{-\gamma s}+(u-vf) e^{-\beta s}$ whenever $\gamma \ne \beta$ and $e^{-\gamma s} (u+\gamma v s)$ otherwise, where $f \equiv \frac{\beta}{\beta-\gamma}$. Finally, the PGF $G$ is recovered by exponentiating $\phi$. We follow the approach of Bokes \cite{bokes_exact_2012, singh_consequences_2012} in evaluating the PGF for $x,y$ around the complex unit circle, interpreting these values as the two-dimensional discrete Fourier transform (DFT), or characteristic function (CF) values, of the original probability distribution, and converting them to the discrete domain by application of the inverse discrete Fourier transform (IDFT). This method has time complexity $O(\mathcal{N} \log \mathcal{N})$, where $\mathcal{N}$ is the state space size, such that $\mathcal{N} = \max n \times \max m$ of interest (\textbf{Figure 1b}). For systems with relatively low copy numbers up to $\approx 100$, where CME modeling is necessary, $N \sim 100 \times 100$, requiring on the order of 10,000 evaluations of the integral $\int_0^t [M(U(s))-1]ds$.

The model with a geometric burst size distribution of mean $b$ requires the evaluation of $\int_0^t \frac{bU}{1-bU}ds$. This integral does not have a closed-form solution, and must be treated using repeated numerical quadrature. However, an approximation to the integral can be computed by decomposing the integrand into an integrable power series. Any expression in the form of $\frac{X}{1-X}$ is amenable to an expansion in powers of $X=bU$. In the region $|X|>1$, the Laurent expansion $-\sum_{i=0}^\infty X^{-i}$  is available. The intuitive choice of the complementary Taylor expansion $\sum_{i=0}^\infty X^{i}$ , which is valid for $|X|<1$, is inappropriate for integration across the boundary $|X|=1$: the approximation diverges and the integral of the expansion ceases to be identical to the original integral. Instead, we leverage the form of $U$, and note that $Re(U)<0$ for all nontrivial choices of $u,v$. Therefore, we utilize the Taylor expansion about $-1$, which is valid for $|X+1|<2$; the form of the series is $-\sum_{i=0}^\infty 2^{-i-1} (1+X)^{i} - 1/2$. As shown in the illustration of their shared domain of convergence (\textbf{Figure 1c}), it is possible to select the appropriate approximation based solely on a threshold for the real-valued $|U|$, which simplifies the computation.

Thus, $X$ is decomposed into multiple approximation domains $\{S_j\}$, such that $|X|$ evaluated at the boundary $\partial S_j$ is $\alpha$, the threshold choice, and successive domains alternate in having $|X|$ strictly greater or less than $\alpha$ (\textbf{Figure 1d}). As discussed in the \textbf{Supplementary Note}, the form of $U$ guarantees that $|\{S_j\}|\le4$; at most two Laurent and two Taylor approximations are necessary. Examination of the expansions shows that both can be expressed as $\sum_i \Omega_{j,i} U^i$. If $|U(s)|\ge \alpha \forall s\in S_j$, the Laurent approximation is appropriate, and $\Omega_{j,i}=-b^{-i}$. For a Laurent order of approximation $N_L, i\in \{0,-1,-2, ... ,-N_L\}$. Conversely, if  $|U(s)|\le \alpha \forall s\in S_j$, the Taylor approximation is appropriate. For a Taylor order of approximation $N_T$, binomial expansion of $(1+X)^i$ yields $\Omega_{j,i} = \sum_{k=i}^{N_T} b^i 2^{-k-1} \binom{k}{i}$. The resulting approximation $\sum_i \Omega_{j,i}  U^i$ has $i\in \{1,2, ... ,N_T\}$. 

Finally, the full integrand $\frac{bU}{1-bU}$ is approximately $\sum_j \sum_i \Omega_{j,i}  U^i$. Therefore, the sought integral $\int_0^t \frac{bU}{1-bU}ds$ can be computed using the truncated power series  $\sum_j \sum_i \Omega_{j,i}  \int_{S_j} U^i ds$, where each expansion is only integrated over its appropriate domain of convergence $S_j$. The details of computation are provided in the \textbf{Supplementary Note}, and the integrals $\int U^i ds$ are given in \textbf{Table I}. Numerical routines to evaluate the exponential integral and the Gaussian hypergeometric function are readily available; however, they are not necessarily optimized for speed. We discuss the approximation schema used to make them practical for large-scale computation in the \textbf{Supplementary Note}.

Furthermore, the same approach can be used for other burst distributions. We consider a degenerate distribution (a gene locus that produces $b$ transcripts per burst), a uniform distribution (a gene locus equally probably to produce any number of transcripts between $a$ and $b$) \cite{beer_effect_2016, kuwahara_beyond_2015}, and a shifted geometric distribution (a gene locus guaranteed to produce at least one transcript per burst, e.g. due to the inhibitor being removed by an advancing RNA polymerase). We find that the approximate solutions to these systems can also be expressed in the form  $\sum_j \sum_i \Omega_{j,i}  \int_{S_j} U^i ds$, as shown in \textbf{Table II}. Equivalently, as long as numerical routines are available to compute  $\int U^i ds$ for $i\in \ZZ$, a broad array of burst distributions can be computed simply by determining the appropriate integration limits (domains where the expansions converge) and computing the coefficients $\Omega_{j,i}$.

\section{Results and Discussion}

We have presented an approximation for the Chemical Master Equation solution of bursty pre-mRNA production and its conversion to mature mRNA. We explored several burst distributions discussed in previous studies and explored an extension to a polymerase-inhibitor interaction model. The CME solutions can be found via the computation of $ \sum_i \Omega_{i}  \int U^i ds$ for the finite-support distributions and $\sum_j \sum_i \Omega_{j,i}  \int_{S_j} U^i ds$ for the infinite-support distributions. The analytical solutions of $\int U^i ds$ are given in \textbf{Table I}, whereas the combinatorial weights for specific burst distributions are given in \textbf{Table II}. 

The series form of the solution enables the modulation of approximation order for computational facility (\textbf{Figure 2a}). The control of method precision and runtime motivates the development of adaptive methods that determine a broad parameter domain using a low-fidelity approximation, then refine it using higher-order or quadrature-based methods. 

The purpose of the current investigation is the development of a unified framework for the computation of CME solutions for a variety of burst models, as well as the determination of analytical solutions for the approximations. To maintain generality, we do not emphasize a particular implementation of the underlying special functions, but presuppose the availability of efficient implementations of the incomplete gamma and Gaussian hypergeometric function. Nevertheless, as a proof of concept, we develop a case study to benchmark the performance of the degenerate case $\beta=\gamma$. Furthermore, we discuss several considerations for implementation and evaluation of the special functions (\textbf{Supplementary note}). 

In light of the motivating broader goal of parameter estimation, we use the algorithm to compute likelihood (Kullback-Leibler divergence) landscapes for simulated data \cite{gillespie_general_1976} with a geometric burst size distribution and $b=19$, $k_i=2.5$, $\beta=\gamma=1$ (\textbf{Figure 2b}). The landscapes produced by the approximation method (shown for $N_L=N_T=7$) closely follow those produced via numerical integration. Repeating this analysis for a range of approximation orders allows benchmarking the method. Over the entire domain shown in \textbf{Figure 2b}, the quality of approximation can be easily controlled by modulating the Taylor approximation order (\textbf{Figure 2c}). The runtime is largely a function of the Laurent approximation order (\textbf{Figure 2d}), due to its explicit reliance on the computation of special functions. We particularly note that the commercial adaptive quadrature method used for benchmarking \cite{noauthor_matlab_2019} provides poor control of runtime.

The procedure for regenerating the discrete distributions from generating functions presents certain problems for inference. As shown in \textbf{Figure 2a}, the result of the IDFT is not guaranteed to be a probability distribution; the IDFT enforces $\sum_k \pi_k = 1$, but does not enforce $\pi_k \ge 0 \forall k$. However, the properties of the Markov chain ostensibly guarantee that $\pi_k \ge 0$, with the inequality becoming strict at equilibrium. For the computation of divergence, we treat this problem in an ad hoc manner, by setting $\pi_k\le0$ to a small float near machine epsilon. A natural, and potentially valuable, extension of this method is the development of transformations using non-negative, non-Fourier basis functions.

An alternative approach is available, and yields faster performance at the expense of interpretability in the Bayesian framework. Instead of computing Kullback-Leibler divergence in the probability domain, it is possible to compute a measure of distance between the characteristic functions of proposed and observed distributions, or even their corresponding cumulants (logarithms). This approach provides two advantages. Firstly, the roundoff and computational expense of repeated exponentiation and logarithm operations is eliminated. Secondly, the overall computational complexity of an inference procedure that uses the Fourier transform method and samples $\mathcal{M}$ candidate parameters is $O(\mathcal{MN} \log \mathcal{N})$. Performing the entire analysis in the Fourier domain requires only a single Fourier transform to determine the empirical characteristic function, reducing the computational complexity to $O(\mathcal{N} \log \mathcal{N} + \mathcal{MN})$, equivalent to $O(\mathcal{MN})$ in the practical limit of large $\mathcal{M}$. This approach is has been explored for use in goodness-of-fit testing and model selection \cite{lee_inferential_2019,jimenez-gamero_fourier_2016}, but only rarely for parameter inference \cite{bee_characteristic_2018}. However, we anticipate that the computational advantages may outweigh the incompatibility with Bayesian inference, similarly to the recent interest in using nonparametric Kolmogorov and Wasserstein distances for parameter inference \cite{sommerfeld_inference_2018, bernton_parameter_2019, gyorfi_minimum_1996}. Further, we note that optimization of the characteristic function uses information about the entire distribution, potentially overcoming identifiability issues observed using other computationally inexpensive non-Bayesian approaches, such as the method of moments \cite{munsky_distribution_2018}. Since characteristic function methods have primarily been used for analysis of (continuous) stable distributions \cite{yu_empirical_2004, feuerverger_efficiency_1981, bee_characteristic_2018} and their performance for inference from discrete-valued random variable observations has not, to our knowledge, been systematically explored, signifying a substantial lacuna. Thus, this approach is a natural next step for optimizing inference from large datasets.

Our discussion of parameter estimation only touched upon inference from steady-state data, which is relevant for fixed-cell experiments that produce information about molecule distributions without a natural time coordinate, such as those available via scRNA-seq \cite{kolodziejczyk_technology_2015} and smFISH \cite{femino_visualization_1998, raj_imaging_2008}. However, experimental methods with temporal information are available \cite{golding_real-time_2005, garcia_quantitative_2013, specht_critical_2017, schofield_timelapse-seq:_2018, herzog_thiol-linked_2017}. Given live-cell data, where cell identities are tracked across time, it is straightforward to extend this method to compute the probability of transitioning from an initial state to any other state, and thus compute the full likelihood of a time-series (\textbf{Supplementary Note: Addenda}). Repeating this process for all observed cells, assuming their trajectories are independent, and summing the log-likelihoods of their time-series yields a joint likelihood for the observations of the entire experiment \cite{daigle_accelerated_2012, corrigan_continuum_2016, golightly_bayesian_2011}. Furthermore, given fixed-cell data, where only the \textit{population-level statistics} are tracked across time, it is likewise straightforward to compute the probability of transitioning from one copy-number distribution to another, and use it for likelihood computation \cite{poovathingal_global_2010} (\textbf{Supplementary Note: Addenda}).

Finally, technical challenges in single-cell transcriptomics, such as sparsity of sampling in sequencing \cite{grun_validation_2014} and noise in fluorescence microscopy \cite{sgouralis_introduction_2017}, have resulted in alternative competing explanations for qualitative features of observed biomolecule distributions, such as heavy-tailed laws \cite{ham_extrinsic_2019, singh_consequences_2012} and apparent dropouts \cite{qiu_embracing_2018, svensson_droplet_2019, andrews_false_2019, li_accurate_2018}. We anticipate that intrinsic degeneracies, as well as aleatory effects, in mapping from a model parameter space to an observable space preclude the unambiguous identification of underlying biophysical schema: the presence of parameter equivalence classes, even in inference of simple models, is well-characterized \cite{weinreb_fundamental_2018, weber_identification_2018, cinquemani_identifiability_2018, gorin_stochastic_2020}. Nevertheless, we also anticipate that the development of analytical solutions, as well as numerical solvers, for a diversity of transcriptional mechanisms, sampling behaviors, and multimodal observables will aid in making inference sufficiently robust for design and extrapolation. For example, as a natural extension, it is straightforward to calculate the laws for observed pre-mRNA and mRNA copy numbers by computing the distributions under an arbitrary sampling schema. This approach enables the natural integration of experimental noise in the same framework as the underlying transcriptional and molecular stochasticity, enabling the simultaneous inference of experimental and physiological parameters.

\section{Acknowledgments}

The DNA, pre-mRNA, and mature mRNA used in \textbf{Figure 1a} are derivatives of the DNA Twemoji by Twitter, Inc., used under CC-BY 4.0. The routine for computing the Taylor approximation coefficient $\Omega_{j,i}$ uses a function by Ben Barrowes \cite{barrowes_computation_2005}, translated from the FORTRAN original by Zhang and Jin \cite{zhang_computation_1996}. The routine for computing the Taylor series approximation to the exponential integral $E_1 (z)$ is a heavily modified version of a function by Ben Barrowes  \cite{barrowes_computation_2005}, translated from the FORTRAN original by Zhang and Jin \cite{zhang_computation_1996}. G.G. and L.P. were partially funded by NIH U19MH114830.

\section{Data and Code Availability}

The algorithm for the degenerate system is available at \verb|https://github.com/pachterlab/GP_2020|, along with MATLAB codes to reproduce \textbf{Figure 2} and \textbf{Supplementary Figure 1}.

\addcontentsline{toc}{section}{References}

\bibliography{references}

\pagebreak
\section{Figures}
\begin{figure}[h]
\centering
\includegraphics[width=0.7\columnwidth]{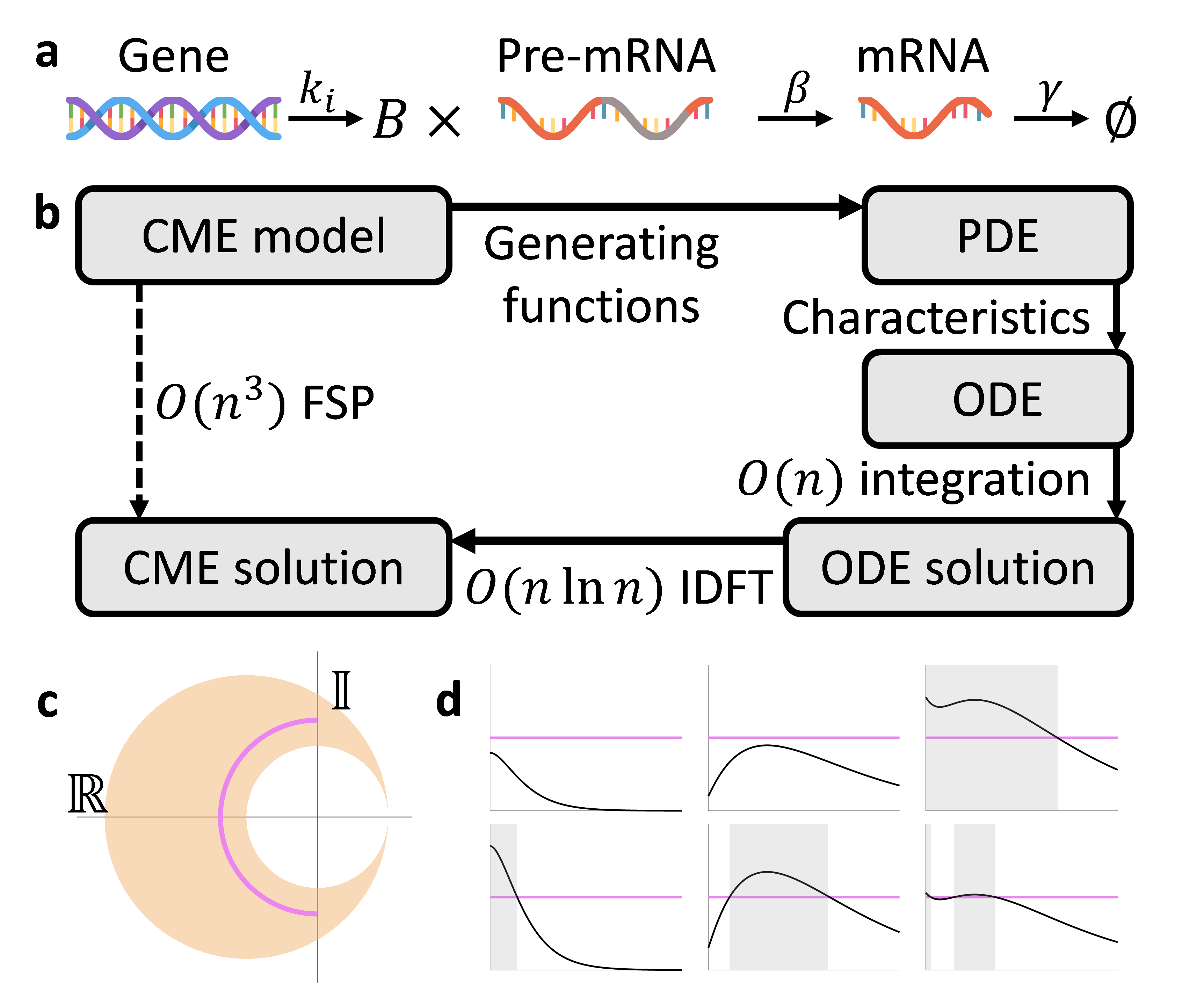}
\caption{(a) Schema of modeled physiology ($k_i$: burst frequency; $B$: burst size drawn from discrete distribution on $\NN$; $\beta$: pre-mRNA splicing rate; $\gamma$: mRNA degradation rate). (b) Outline of the solution procedure. (c) Taylor/Laurent approximation criterion (orange: approximations’ common region of convergence; purple: threshold value of $|U|$). (d) Sample shapes of $|U|$ and their partitions (black curve: $|U|$; purple: threshold value of $|U|$; grey: Laurent approximation regions).}
\label{fig:fig1}
\end{figure}

\begin{figure}[p]
\centering
\includegraphics[width=0.7\columnwidth]{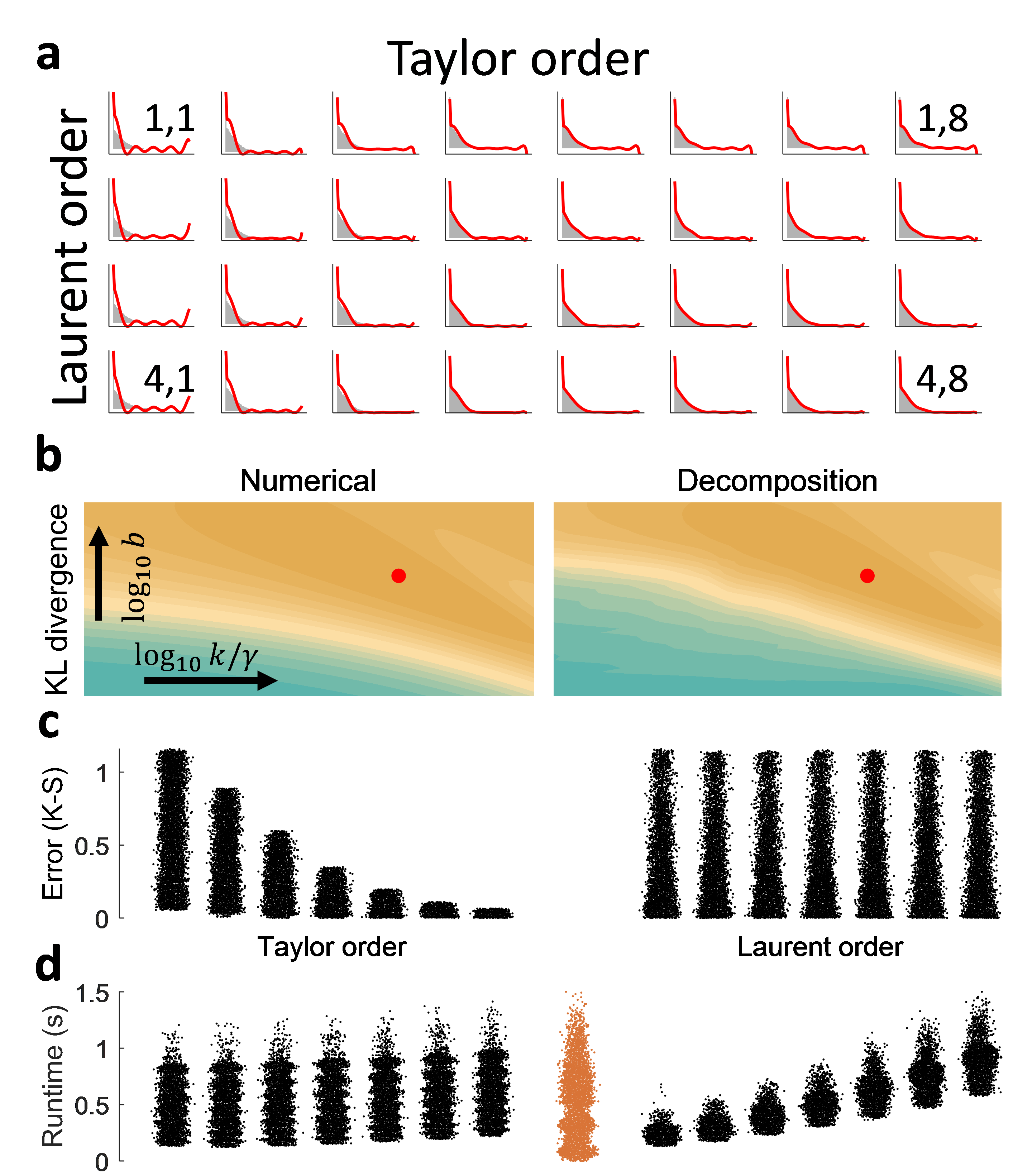}
\caption{(a) Comparison of marginal mature mRNA copy-number distributions for a range of approximation orders ($\#,\#$ tuple and subplot location: Laurent, Taylor approximation order; grey: histogram from $10^5$ Gillespie simulations; red line: distribution calculated from approximation). 
(b) Likelihood landscape for a set of simulated steady-state data with $\gamma=\beta$, calculated over $50 \times 50$ trial parameter combinations (numerical: quadrature-based computation; decomposition: expansion-based computation; abscissa: $\log_{10} k_i / \gamma \in [-1,1]$; ordinate: $\log_{10} b \in (0,2]$; orange: low divergence; teal: high divergence, red point: ground truth). (c) Kolmogorov-Smirnov error between quadrature- and expansion-based joint distributions for parameter sets in (b), calculated for combinations of Taylor and Laurent orders $\in\{1,...,7\}\times\{1,...,7\}$ (black point: single parameter set; uniform jitter added). (d) Joint distribution calculation times, determined over the domain in (b) and approximation orders in (c) (black point: single parameter set computed using expansions; orange point: single parameter set computed using numerical quadrature; uniform jitter added). 
}
\label{fig:fig2}
\end{figure}

\pagebreak

\section{Tables}

\begin{center}
 \begin{tabular}{|c | c |c |c|} 
 \hline
   &   & $\gamma=\beta$ & $\gamma \ne \beta$ \\ 
 \hline
 $U$ &   & $e^{-\gamma s} (u+\gamma vs)$ & $U(s)=vfe^{-\gamma s}+(u-vf) e^{-\beta s}$ \\ 
 \hline
 Taylor & $v=0$ & $-\frac{u^i}{i\gamma} e^{-i \gamma s}$ & $-\frac{u^i}{i\beta} e^{-i \beta s}$ \\
 \hline
   & $v\ne 0$ & 
   $-\frac{1}{\gamma i} (\frac{v}{i})^i i! \sum_{j=0}^i \frac{i^j}{j!} \bigg[(\frac{u}{v}+\gamma s)^j e^{-i \gamma s} \bigg]$ 
   & 
   \makecell{$\frac{1}{\beta-\gamma} (u-vf)^i$ \\ $\times \sum_{j=0}^i \binom{i}{j} (\frac{vf}{u-vf})^j \frac{1}{j-if} e^{[j(\beta-\gamma)-i\beta]s}$}    \\
 \hline
 Laurent & $v=0$ & $\frac{u^{-i}}{i\gamma} e^{i \gamma s}$ & $\frac{u^{-i}}{i\beta} e^{i \beta s}$ \\
 \hline
  & $v\ne0$ & $-\frac{e^{-iu/v}}{\gamma v} (u+\gamma vs)^{1-i} E_i \bigg(-\frac{i}{v}(u+\gamma vs)\bigg)$ &
  \makecell{ $\frac{(vf-u)^\rho (vf)^{-i-\rho}}{(\gamma-\beta) \rho \big[\frac{u-vf}{vf} e^{-(\beta-\gamma)s} \big]^{\rho}}$
  \\ 
  $\times {}_2 F_1 (i,-\rho;-\rho+1;\frac{u-vf}{vf}e^{-(\beta-\gamma)s})$}
  \\ 
 \hline
\end{tabular}
\end{center}
\textbf{Table I}. Integrals of $U^i$ for various approximations and levels of degeneration.
\begin{center}
 \begin{tabular}{|c | c |c |c|c|} 
 \hline 
 &   \multicolumn{4}{|c|}{Burst distribution} \\
 \hline
 & $b$-step & Uniform & Geometric & Shifted geometric \\
 \hline
 $\frac{1}{k_i} \frac{\partial \phi}{\partial s}$ 
 & $(1-U)^b-1$ & $\frac{1}{n} \sum_{i=a}^b (1+U)^i -1$ 
 & $\frac{bU}{1-bU}$
 & $\frac{bU}{1+(1-b)U}$\\
 \hline
 $\Omega_{j,i}$ 
 & $\binom{b}{i}$
 & $\frac{1}{n}\big[\binom{b+1}{i+1}-\binom{a}{i+1}\big]$
 & \makecell{$b^i \sum_{k=i}^{N_T} \frac{1}{2^{k+1}} \binom{k}{i}$ \\ $-b^{-i}$}
 & \makecell{$b(b-1)^{i-1} \sum_{k=i}^{N_T} \frac{1}{2^{k+1}} \binom{k}{i}$ \\ $-b (b-1)^{-i-1}$}\\
 \hline
 $i$ & $1,...,b$& $1,...,b$& \makecell{$1,...,N_T$ \\$0,-1,...,-N_L$}& \makecell{$1,...,N_T$ \\$0,-1,...,-N_L$}\\
 \hline
 $U$ & $\CC$ & $\CC$ & \makecell{$|U|<\frac{1+\sqrt{3}}{2b}$ \\ $|U|>\frac{1+\sqrt{3}}{2b}$ } & \makecell{$|U|<\frac{1+\sqrt{3}}{2(b-1)}$ \\ $|U|>\frac{1+\sqrt{3}}{2(b-1)}$ }\\
 \hline
\end{tabular}
\end{center}
\textbf{Table II}. Integrands, expansion coefficients, summation indices, and expansion domain thresholds associated with approximating the CME solutions for four burst distributions. 

\end{document}


\title{Special Function Methods for Bursty Models of Transcription Supplementary Note}
\author{Gennady Gorin, Lior Pachter}
\date{ }

\maketitle

\tableofcontents

%

\section{Burst generating functions and their expansions}

Let $\phi$ be the logarithm of the probability generating function (PDF) $G$. 
To determine $\phi$, it is necessary to integrate $\frac{d \phi}{dt} = k_i (M(U(s))-1)$, where $M(U(s))$ is the factorial-moment generating function (FMGF) of the burst distribution. This function is represented as $M(U) = F(1+U)$ where $F(x) = \sum_{\rho=0}^{\infty} P(B=\rho) x^\rho$ is the PGF of the burst distribution. 

In what follows we use $\mu$ to denote the mean burst size for a single-parameter burst distribution.

\subsection{Geometric distribution, $\mu=b$}

As discussed by Singh and Bokes \cite{singh_consequences_2012}, the geometric distribution with mean burst size $b$ has the probability mass function (PMF) $P(B=\rho) = p(1-p)^\rho$, where $p=\frac{1}{1+b}$ and $\rho=0,1,2,...$. The resulting PGF is $F(x) = E[x^\rho] = \sum_{\rho=0}^{\infty} P(B=\rho) x^\rho =  \sum_{i=0}^{\infty} p(1-p)^\rho x^\rho = p\sum_{i=\rho}^{\infty} [(1-p) x]^\rho = p\frac{1}{1-(1-p)x}$. This is exact for $|x|\le 1$ and extends to $x\in \CC$ by analytical continuation. Using the definition of $p$, $F(x) = \frac{1}{1+b} \frac{1}{1-\frac{b}{1+b}x} = \frac{1}{1+b-bx}$. The transformed PGF $M(u) = F(1+u) = \frac{1}{1+b-b(1+u)} = \frac{1}{1-bu}$. Finally, $M(u)-1 = \frac{1}{1-bu} - 1 = \frac{bu}{1-bu}$, which recapitulates previous work.

Defining $X=bu$, we have that $M(u)-1 = \frac{X}{1-X}$. This expression has the well-known Laurent expansion $-\sum_{i=0}^\infty \frac{1}{X^i}$ for all $|X|>1$. Since $b$ is real-valued, $M(u)-1 = \sum_{i=0}^\infty \Omega_i u^{-i}$, where $\Omega_i = -b^{-i}$. This expansion can be truncated at order $N_L$, yielding $M(u)-1 \approx \sum_{i=0}^{N_L} \Omega_i u^{-i}$.

Within the region $|X+1|<2$, the function $\frac{X}{1-X}$ has the Taylor expansion $\sum_{i=1}^\infty \frac{(1+X)^i}{2^{i+1}} - \frac{1}{2}$. Expanding the binomial yields $\sum_{i=1} 2^{-i-1} \sum_{j=0}^i \binom{i}{j} X^i - \frac{1}{2} = \sum_{i=1} \frac{1}{2^{i+1}} \sum_{j=1}^i \binom{i}{j} X^j = \sum_{i=1} \frac{1}{2^{i+1}} \sum_{j=1}^i \binom{i}{j} (bu)^j$. This expansion can be truncated at order $N_T$, yielding $M(u)-1 \approx \sum_{i=1}^{N_T} \frac{b^i}{2} (\sum_{j=i}^{N_T} \frac{1}{2^j} \binom{j}{i}) u^i = \sum_{i=1}^{N_T} \Omega_i u^i$, where $\Omega_i = \frac{b^i}{2} \sum_{j=i}^{N_T} \frac{1}{2^j} \binom{j}{i}$. 

Equivalently, $\Omega_i = b^i[1-2^{-N_T-2} \binom{N_T+1}{i} {}_2 F_1(1,N_T+2; -i+N_T+2; \frac{1}{2})]$.

\subsection{Shifted geometric distribution, $\mu=b$}

A shifted geometric distribution with mean burst size $b$ has the PMF $P(B=\rho) = p(1-p)^{\rho-1}$, where $p=\frac{1}{b}$ and $\rho=1,2,3,...$. The resulting PGF over the relevant support is $F(x) = E[x^\rho] = \sum_{\rho=1}^{\infty} P(B=\rho) x^\rho =  \sum_{\rho=1}^{\infty}p(1-p)^{\rho-1} x^\rho = p\sum_{\rho=0}^{\infty}(1-p)^\rho x^{\rho+1} = px \sum_{\rho=0}^{\infty} [(1-p)x]^\rho = px \frac{1}{1-(1-p)x}$. Using the definition of $p$, $F(x) = \frac{x}{b} \frac{1}{1-\frac{b-1}{b}x} = \frac{x}{b-(b-1)x}$. The transformed PGF $M(u) = F(1+u) = \frac{1+u}{b-(b-1)(1+u)} = \frac{1+u}{b-b+1-bu+u} = \frac{1+u}{1+(1-b)u}$. Finally, $M(u)-1 =\frac{1+u - 1 - (1-b)u}{1+(1-b)u} = \frac{bu}{1+(1-b)u}$.

Within the region $|u| > \frac{1}{|b-1|}$, the function $M(u)-1$ has the Laurent expansion $-\frac{b}{b-1} \sum_{i=0}^\infty \frac{1}{(b-1)^i u^i} = \sum_{i=0}^\infty \Omega_i u^{-i}$, where $\Omega_i = -\frac{b}{(b-1)^{i+1}}$. This expansion can be truncated at order $N_L$, yielding $M(u)-1 \approx \sum_{i=0}^{N_L} \Omega_i u^{-i}$. 

Defining the intermediate variable $a = \frac{1}{b-1}$ and noting $a+1 = \frac{b}{b-1}$, $M(u)-1 = \frac{b}{b-1} \frac{(b-1)u}{1+(1-b)u} = - \frac{b}{b-1} \frac{(1-b)u}{1+(1-b)u} = - \frac{b}{b-1} \frac{u}{u - \frac{1}{1-b}} =  - \frac{b}{b-1} \frac{u}{u-a}$. The Taylor expansion of $\frac{u}{u-a}$ about $-a$, within the domain $|u+a|<2|a|$, is $-\sum_{i=1}^\infty \frac{1}{2^{i+1} a^i} (a+u)^i + \frac{1}{2} = -\sum_{i=1}^\infty \frac{1}{2^{i+1} a^i} \sum_{j=0}^i \binom{i}{j} u^j a^{i-j} + \frac{1}{2}  \\
= -\sum_{i=1}^\infty \frac{1}{2^{i+1}} \sum_{j=0}^i \binom{i}{j} u^j a^{-j} - \frac{1}{2} = -\sum_{i=1}^\infty \frac{1}{2^{i+1}} \sum_{j=1}^i \binom{i}{j} u^j a^{-j}$. Therefore, $M(u)-1 \\
= \frac{b}{2(b-1)} \sum_{i=1}^\infty \frac{1}{2^i} \sum_{j=1}^i \binom{i}{j} u^j (b-1)^j$. This expansion can be truncated at order $N_T$, yielding $\\M(u)-1 \approx \sum_{i=1}^{N_T} \Omega_i u^i$, where $\Omega_i = \frac{b}{2}(b-1)^{i-1} \sum_{j=i}^{N_T} \frac{1}{2^j} \binom{j}{i}$. 

Equivalently, $\Omega_i = b(b-1)^{i-1} [1-2^{-N_T-2} \binom{N_T+1}{i} {}_2 F_1(1,N_T+2;-i+N_T+2,\frac{1}{2})]$.


\subsection{Degenerate ($b$-step)}

The degenerate burst distribution that yields $b$ pre-mRNA products with every burst has PMF $P(B=\rho) = \delta_{\rho b}$, where $\delta_{ij}$ is the Kronecker delta. The resulting PGF is  $F(x) = E[x^\rho] = \sum_{\rho=0}^{\infty} P(B=\rho) x^\rho = x^b$. Therefore, $M(u)-1 = (1+u)^b-1$.

Expanding the binomial yields the expression $M(u)-1 = \sum_{i=0}^b \binom{b}{i} u^i - 1 = \sum_{i=1}^b \binom{b}{i} u^i = \sum_{i=1}^b \Omega_i u^i$, where $\Omega_i = \binom{b}{i}$.

\subsection{Uniform on $[a,b]$}

The uniform distribution on $[a,b]$ has PMF $P(B=\rho) = \frac{1}{n} I(\rho\in [a,b])$, where $I(\cdot)$ is the indicator function and $n=b-a+1$. The resulting PGF over the relevant support is $F(x) = E[x^\rho] = \frac{1}{n} \sum_{\rho=a}^b x^\rho$. Therefore, $M(u)-1 = \frac{1}{n} \sum_{i=a}^b (1+u)^i-1$. 

Expanding the binomial yields $M(u)-1 = \frac{1}{n} \sum_{i=a}^b \sum_{j=0}^i \binom{i}{j} u^j - 1 = \frac{1}{n} \sum_{i=a}^b \sum_{j=1}^i \binom{i}{j} u^j$. Assuming $a>0$, reversing the order of summation yields $M(u)-1 =  \sum_{j=1}^b \frac{W_1(j)-W_2(j)}{n(j+1)} u^j$, where $W_1(j) = \frac{\Gamma(b+2)}{\Gamma(b-j+1)}$ for all $j$ and $W_2(j) = 0$ for $j\in [a,b]$ and $\frac{\Gamma(a+1)}{\Gamma(a-j)}$ otherwise. Therefore, $M(u)-1 = \sum_{i=1}^b \Omega_i u^i$, where $\Omega_i = \frac{W_1(i)-W_2(i)}{n(i+1)}$. Equivalently, $\Omega_i = \frac{1}{n} [\binom{b+1}{i+1} - \binom{a}{i+1}]$.

\section{Special function solutions to $\int U^i ds$}

As explored in the section "\textbf{Burst generating functions and their expansions}," the FCGFs of the burst models explored here can be represented in the common form $M(u)-1 \approx \sum_{i} \Omega_i u^i$, where $i \in \ZZ$. Given $U(s)$, a characteristic solution representing the dynamics downstream of the gene locus, the integral $\frac{\phi}{k_i} = \int [M(U(s))-1] ds$ can be approximated as $\sum_i \Omega_i \int U(s)^i ds$ wherever the expansion holds. The determination of the appropriate domains is treated in the section "\textbf{Domain Decomposition}."

\subsection{Degenerate case: $U(s; u,v) = e^{-\gamma s} (u+\gamma vs)$}

\subsubsection{Taylor expansion}

From standard identities \cite{abramowitz_handbook_1970}, the Taylor expansion of order $i\in \ZZ$ yields $T_i(s;u,v) = \int e^{-i\gamma s} (u+\gamma vs)^i ds = -\frac{e^{iu/v} v^i}{\gamma i^{i+1}} \Gamma(i+1, \frac{i}{v}(u+\gamma vs))$, where $\Gamma(a,z) = \int_z^\infty t^{a-1} e^{-t} dt$ is the upper incomplete Gamma function \cite{abramowitz_handbook_1970}. The evaluation of $\Gamma(a,z)$ for arbitrary complex arguments is computationally nontrivial. However, for $n\in \NN$, $\Gamma(i+1,z) = i! e^{-z} e_n(z)$, where $e_n(z) = \sum_{k=0}^n \frac{z^k}{k!}$. 

Therefore, $T_i = -\frac{e^{iu/v} v^i}{\gamma i^{i+1}} i! e^{-\frac{i}{v}(u+\gamma vs)} \sum_{k=0}^i \frac{1}{k!}(\frac{i}{v}(u+\gamma vs))^k =-\frac{e^{- i\gamma s}v^i}{\gamma i^{i+1}} i! \sum_{k=0}^i \frac{i^k}{k!}(\frac{u}{v} + \gamma s)^k$. The resulting definite integral from $s_1$ to $s_2$ is simply $-\frac{v^i i!}{\gamma i^{i+1}} \sum_{k=0}^i \frac{i^k}{k!}[(\frac{u}{v} + \gamma s_2)^k e^{- i\gamma s_2} - (\frac{u}{v} + \gamma s_1)^k e^{- i\gamma s_1}]$, which is directly computable without the use of special function routines.

This expression for $T_i$ is inappropriate for the degenerate case $v=0$, corresponding to the marginal with respect to the pre-mRNA. In that case, the definite integral reduces to $-\frac{u^i}{\gamma i} [e^{-i \gamma s_2} - e^{-i\gamma s_1}]$.

\subsubsection{Laurent expansion}

For $k=-i$, where the order of Laurent expansion $k\in \ZZ$, the identity $L_k(s;u,v) \\
=\int e^{i\gamma s} (u+\gamma vs)^{-i}ds = -\frac{e^{-iu/v} v^{-i}}{\gamma (-i)^{1-i}} \Gamma(1-i, -\frac{i}{v}(u+\gamma vs))$ holds.

$L_k = \frac{e^{-iu/v} (-1)^i}{\gamma v^i i^{1-i}} \Gamma(1-i, -\frac{i}{v}(u+\gamma vs))$. The direct computation of $\Gamma(1-i,z)$ is computationally nontrivial, and finite power series expansions are unavailable for this purpose. However, the following identity holds:

$\Gamma(1-i,z) = z^{1-i} E_i(z)$, where $E_i(z)$ is the generalized exponential integral of order $i$. Further, $E_i(z) = \frac{(-z)^{i-1}}{(i-1)!} E_1(z) + \frac{e^{-z}}{(i-1)!} \sum_{k=0}^{i-2} (i-k-2)! (-z)^k$, where $E_1(z) = \int_z^\infty \frac{e^{-t}}{t} dt$ is the first-order complex exponential integral.

Therefore, $L_k = \frac{e^{-iu/v} (-1)^i}{\gamma v^i i^{1-i}} [-\frac{i}{v}(u+\gamma vs)]^{1-i} E_i(-\frac{i}{v}(u+\gamma vs))\\
= - \frac{e^{-iu/v}}{\gamma v} [u+\gamma vs]^{1-i} E_i(-\frac{i}{v}(u+\gamma vs))$. Using the relation for $E_i(z)$,

$E_i(-\frac{i}{v}(u+\gamma vs)) = \frac{1}{\Gamma(i)} [\frac{i}{v}(u+\gamma vs)]^{i-1} E_1(z) + \frac{1}{\Gamma(i)} \exp(\frac{i}{v}(u+\gamma vs)) \sum_{k=0}^{i-2} (i-k-2)! (\frac{i}{v}(u+\gamma vs))^k$, where $z=-\frac{i}{v}(u+\gamma vs)$.

This yields $L_k = - \frac{e^{-iu/v}}{\gamma v \Gamma(i)} (\frac{i}{v})^{i-1} E_1(z) - \frac{e^{i \gamma s}}{\gamma v \Gamma(i)}  \sum_{k=0}^{i-2} \Gamma(i-k-1) (\frac{i}{v})^k (u+\gamma vs)^{k+1-i}$. 
Finally, the definite integral $L_k(s_2;u,v) - L_k(s_1;u,v)$ is 
\\ $- \frac{e^{-iu/v}}{\gamma v \Gamma(i)} (\frac{i}{v})^{i-1} [E_1(z_2)-E_1(z_1)] \\
- \frac{1}{\gamma v \Gamma(i)}  \sum_{k=0}^{i-2} \Gamma(i-k-1) (\frac{i}{v})^k [(u+\gamma vs_2)^{k+1-i} e^{i \gamma s_2} - (u+\gamma v s_1)^{k+1-i} e^{i \gamma s_1}]$,
where $z_{l} = -\frac{i}{v}(u+\gamma vs_l)$.  
Equivalently, this is \\
$- \frac{e^{-iu/v}}{\gamma v \Gamma(i)} (\frac{i}{v})^{i-1} [E_1(z_2)-E_1(z_1)]\\ - \frac{e^{i\gamma s_1}}{\gamma v \Gamma(i)}  \sum_{k=0}^{i-2} \Gamma(i-k-1) (\frac{i}{v})^k [(u+\gamma vs_2)^{k+1-i} e^{i \gamma (s_2-s_1)} - (u+\gamma v s_1)^{k+1-i}]$.

As in the Taylor expansion, this expression is inappropriate for the degenerate case $v=0$. In that case, the definite integral reduces to $\frac{1}{i u^i \gamma} (e^{i\gamma s_2} - e^{i \gamma s_1})$.

\subsection{Non-degenerate case: $U(s; u,v) = vfe^{-\gamma s} + (u-vf) e^{-\beta s}$}

As throughout, $f$ is defined as $\frac{\beta}{\beta-\gamma}$. 

\subsubsection{Taylor expansion}

For a Taylor expansion of order $i\in \ZZ$, $T_i(s;u,v) = \int [vfe^{-\gamma s} + (u-vf) e^{-\beta s}]^i ds \\
= \int (A e^{-\gamma s} + B e^{-\beta s})^i ds$. Expanding the binomial term yields $T_i 
\\ = \int \sum_{j=0}^i \binom{i}{j} (A e^{-\gamma s})^j (B e^{-\beta s})^{i-j} ds 
= \int \sum_{j=0}^i \binom{i}{j} A^j B^{i-j} e^{-j \gamma s} e^{-(i-j) \beta s} ds \\
= \int \sum_{j=0}^i \binom{i}{j} A^j B^{i-j} e^{[-j \gamma -(i-j) \beta] s} ds =
\sum_{j=0}^i \binom{i}{j} A^j B^{i-j} \int e^{[-j \gamma - (i-j) \beta] s} ds 
\\ = \sum_{j=0}^i \binom{i}{j} A^j B^{i-j} \frac{1}{-j \gamma -(i-j) \beta} e^{[-j \gamma -(i-j) \beta] s} = \sum_{j=0}^i \binom{i}{j} (vf)^j (u-vf)^{i-j} \frac{1}{-j \gamma -(i-j) \beta} e^{[-j \gamma -(i-j) \beta] s}$. \\ The resulting definite integral, which can be computed directly, is: \\
$\sum_{j=0}^i \binom{i}{j} (vf)^j (u-vf)^{i-j} \frac{1}{-j \gamma -(i-j) \beta} [e^{[-j \gamma -(i-j) \beta] s_2}-e^{[-j \gamma -(i-j) \beta] s_1}] \\
= (u-vf)^i \sum_{j=0}^i \binom{i}{j} (\frac{vf}{u-vf})^j \frac{1}{j (\beta - \gamma) -i \beta} [e^{[j (\beta - \gamma) -i \beta] s_2}-e^{[j (\beta - \gamma) -i \beta] s_1}]$. 
\\$= \frac{1}{\beta-\gamma} (u-vf)^i \sum_{j=0}^i \binom{i}{j} (\frac{vf}{u-vf})^j \frac{1}{j - i f} [e^{[j (\beta - \gamma) -i \beta] s_2}-e^{[j (\beta - \gamma) -i \beta] s_1}]$

In the degenerate case $v=0$, the antiderivative is simply $\int ([ue^{-\beta s}]^i ds = u^i \int e^{- i \beta s} ds = -\frac{u^i}{i\beta} e^{-i\beta s}$, with the definite integral $-\frac{u^i}{i\beta} [e^{-i\beta s_2}-e^{-i\beta s_1}]$.

\subsubsection{Laurent expansion}

For a Laurent expansion of order $i\in \ZZ$, $L_i(s;u,v) = \int [vfe^{-\gamma s} + (u-vf) e^{-\beta s}]^{-i} ds \\
= \int (A e^{-\gamma s} + B e^{-\beta s})^{-i} ds$. Considering the integrand:

$l(s;u,v) = (A e^{-\gamma s} + B e^{-\beta s})^{-i} = \frac{1}{(A e^{-\gamma s} + B e^{-\beta s})^i} = \frac{1}{(A e^{-\gamma s} + B e^{-\beta s})^i} = \frac{1}{(1 + \frac{B}{A} e^{-(\beta -\gamma) s})^i A^i e^{- i \gamma s}}$. Defining $z=-\frac{B}{A} e^{-(\beta -\gamma) s}$, this yields $l=\frac{1}{(1-z)^i A^i e^{- i \gamma s}}$. Furthermore, $e^{- i \gamma s} = [e^{-(\beta -\gamma) s}]^\rho$ for $\rho = \frac{-(\beta-\gamma)}{-i \gamma} = \frac{\beta-\gamma}{i \gamma}$. Therefore, $z^\rho = (-\frac{B}{A} e^{-(\beta -\gamma) s})^{\rho} = (-\frac{B}{A})^\rho e^{-i\gamma s}$ and $A^i e^{- i \gamma s} = A^i (-\frac{B}{A})^{-\rho} z^\rho$. This, in turn, yields $l = (-\frac{B}{A})^\rho A^{-i} (1-z)^{-i} z^{-\rho}$. 

Using variable substitution, $\int_{s_1}^{s_2} l(s;u,v) ds  = \int_{z_1}^{z_2} l(z;u,v) dz$, where $z_i = -\frac{B}{A} e^{-(\beta -\gamma) s_i}$. $\frac{dz}{ds} = -(\beta-\gamma)z$, yielding $dz = -(\beta-\gamma)^{-1} z^{-1} ds$. 

The solution to $\int_{z_1}^{z_2} l(s;u,v) dz$ is given by the incomplete beta function $B(z;\mu_1,\mu_2)$ \cite{abramowitz_handbook_1970}:

$r= (-\frac{B}{A})^\rho A^{-i} (\gamma-\beta)^{-1} [B(z_2;\mu_1,\mu_2)-B(z_1;\mu_1,\mu_2)]$, , $\mu_1=-\rho$, and $\mu_2=1-i$. This solution is, in principle, exact. However, methods for the computation of $B(z;\mu_1,\mu_2)$ for arbitrary complex arguments $z$ are unavailable \cite{abramowitz_handbook_1970}, presumably due to the ubiquity of the function in the field of statistical computation, which only requires evaluation of $B(x;\mu_1,\mu_2)$ for $x$ on the real line. Therefore, it is more practical to use the Gaussian hypergeometric function representation $B(z;\mu_1,\mu_2) \\
= \frac{z^{\mu_1}}{\mu_1} {}_2 F_1 (1-\mu_2,\mu_1;\mu_1+1;z)$, yielding the desired result:

$L_i(s;u,v) =  (-\frac{B}{A})^\rho A^{-i} (\gamma-\beta)^{-1} \rho^{-1} [z_2^{-\rho} {}_2 F_1 (i,-\rho;-\rho+1;z_2) - z_1^{-\rho} {}_2 F_1 (i,-\rho;-\rho+1;z_1)]$. 

In the degenerate case $v=0$, the antiderivative is simply $\int ([ue^{-\beta s}]^{-i} ds = u^{-i} \int e^{i \beta s} ds = \frac{1}{i\beta u^i} e^{i\beta s}$, with the definite integral $ \frac{1}{i\beta u^i} [e^{i\beta s_2}-e^{i\beta s_1}]$.

If a robust evaluation routine for ${}_2 F_1$ is available, it is also straightforward to compute the Taylor terms $T_i(s;u,v) = L_{-i}(s;u,v)$ without explicitly performing the summation described above.

\section{Considerations for the numerical evaluation of $\phi(u,v)$}

\subsection{Numerical stability}

\subsubsection{Binomial coefficients}

The expressions for $\Omega_i$ described in the section "\textbf{Burst generating functions and their expansions}" require the computation of binomial coefficients. For the geometric and shifted geometric distributions of burst sizes, which have binomial coefficients on the order of $\binom{N_T}{\cdot}$ (for small integer $N_T$), calculating the coefficients \textit{via} the log-gamma function $\ln \Gamma$ is recommended, per $\binom{m}{n} = \exp(\ln \Gamma(m+1) - \ln \Gamma(n+1) - \ln \Gamma(m-n+1))$. For the degenerate and uniform distributions (with modest to large $b$), calculating the coefficients \textit{via} $\ln \Gamma$ is required for practical implementation.

\subsubsection{Overflow and underflow for large $b$}

Although the approximation $\sum_i \Omega_i \int U(s)^i ds$ converges to the desired generating function $\frac{\phi}{k_i}$, the series as written is not ideal for numerical evaluation by the independent computation of $\Omega_i$ and $\int U(s)^i ds$. In particular, the Taylor coefficients for the geometric and shifted geometric distribution scale with $b^i$ and $(b-1)^i$ respectively. Similarly, their Laurent coefficients scale with  $b^{-i}$ and $(b-1)^{-i}$. Since $b$ may be fairly large (up to 300) for mammalian systems \cite{dar_transcriptional_2012}, the computation of its powers can produce overflow or underflow problems for Taylor and Laurent coefficients, respectively.

Conversely, the Taylor integrals for the degenerate system scale with $\frac{1}{i^i}$, and the Laurent ones scale with $i^i$. 

In case of the degenerate  system, issues can be ameliorated somewhat by computing $\frac{\Omega_i}{i^i}$ and $i^i \int U(s)^i ds$, which yields a $(\frac{b}{i})^i$ term and balances the growth behavior. 

It is not clear whether an analogous approach is available for the non-degenerate case, although solutions based on Stirling's approximation ($i! \sim  i^i  \sqrt{2\pi i} e^{-i}$) may be viable for larger $i$.

\subsection{Special function evaluation}

\subsubsection{Exponential integral}

The solution for the Laurent expansion of the degenerate system, which only involves the evaluation of a parameter-free exponential integral $E_1(z)$, lends itself to optimization. We found that the following combination of approximations yields no more than $10^{-8}$ relative error with respect to the MATLAB function \verb|expint| throughout the entire complex plane \cite{noauthor_matlab_2019}. In this enumeration, $x \equiv Re(z)$ and $y \equiv Im(z)$. For benchmarking, $\tau = T_M / T_A$, where $T_M$ is the runtime of the built-in function and $T_A$ is the runtime of each approximation routine. The functions were run over a $1000\times 1000$ point grid for $x,y\in [-35,35]$.

\begin{enumerate}
    \item Exterior region $[x/17 + 0.3824)^2 +(y/13)^2 > 1]$: 6th order Pad\'e approximation (Luke p.110-112 \cite{luke_mathematical_1975}, table 4.3): \verb|pade1|. $\tau = 45.7$.
    \item Exterior region $[((x+10)/15)^2 +(y/9.5)^2 > 1]$: 10th order Pad\'e approximation (Luke p.110-112 \cite{luke_special_1969}, table 4.3): \verb|pade2|. $\tau = 40.4$.
    \item Elliptic region $[((x+0.65)/4.05)^2 +(y/4)^2 < 1]$: 10th order Pad\'e approximation (Luke p.107-108 \cite{luke_special_1969}, table 4.2): \verb|pade3|. $\tau = 33.8$.
    \item Elliptic region $[((x+4.5)/4.5)^2 +(y/2.3)^2 < 1]$: 20th order Chebyshev approximation (Luke p.104 \cite{luke_special_1969}, table 4.1): \verb|cheb1|. $\tau = 15.4$.
    \item Radial region $[x<-8 \cap |y| < (-x-8)0.5294)]$: 20th order Chebyshev approximation (Luke p.105 \cite{luke_special_1969}, table 4.1): \verb|cheb2|. $\tau = 12.6$.
    \item Annular region (all other values of $z$): 55th order series approximation (Zhang and Jin \cite{zhang_computation_1996}): \verb|taylor|. $\tau = 7.3$.
\end{enumerate}

The resulting performance is shown in Supplementary Figure $\ref{fig:expint}$. Teal denotes error above $10^{-8}$; brown denotes error below $10^{-8}$; the purple lines are the approximation regions. The combined approximation produced $\tau=24.1$, with maximum error of $10^{-7.9}$. 

\begin{figure}[h]
\centering
\includegraphics[width=\columnwidth]{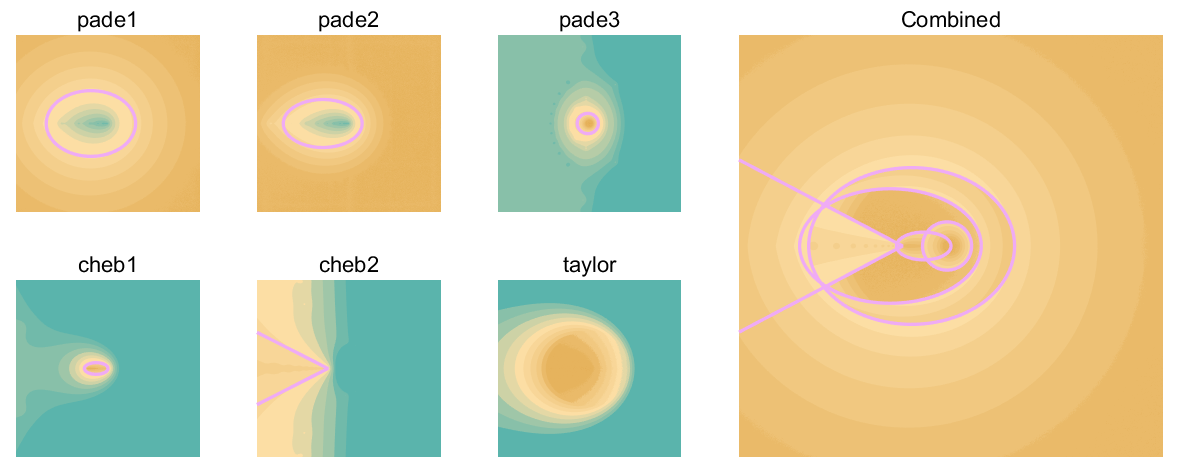}
\caption{Exponential integral approximations.}
\label{fig:expint}
\end{figure}


\section{Domain decomposition: geometrically-distributed bursts}

The Taylor and Laurent approximations are only valid over some domains. For example, the Laurent approximation to $\frac{b\tilde{u}(s)}{1-b\tilde{u}(s)}$ is only valid for $|b\tilde{u}(s)| \equiv |U(s)| > 1$. We split the domain of integration into multiple sub-domains, and integrate each one separately. The domain boundaries are values of $s$ $(s_1, s_2, s_3...)$ where $|U(s)| = \alpha$, such that $\alpha$ is within the domain of convergence for both Taylor and Laurent approximations. 

The number of domains depends upon the number of times the function $|U(s)|$ crosses $\alpha$. From integration from 0 to $t$, where the function crosses $\alpha$ at $s_1, s_2, s_3..., s_i$, the number of domains is $i+1$. By the mean value theorem, for differentiable $|U(s)|$, $i \le j+1$, where $j$ is the number of points such that $\frac{d|U(s)|}{ds}=0$. 

In the present section, we demonstrate that \textit{at most} four domains are necessary; equivalently, $\frac{d|U(s)|}{ds} = 0$ for at most \textit{two} $s\ge 0$.

\subsection{Selection of threshold $\alpha$}

For the geometric distribution of burst sizes, the region of convergence for the Taylor expansion is a circle of radius 2 centered at -1, while the region of convergence for the Laurent expansion is a circle of radius 1 centered at 0. For the relevant case of $Re(U)\le 0$, the minimum distance between the boundaries of the regions of convergence is achieved at $Re(U)=0$. The corresponding region of convergence is $(i,\sqrt{3}i)$. As shown in \textbf{Figure 1c}, we select $\alpha$ equidistant from the two regions' boundaries, as a minimax strategy. This threshold for $|U|$ corresponds to $\alpha = \frac{1+\sqrt{3}}{2b}$. 

For the shifted geometric distribution, the region of convergence for the Taylor expansion is a circle of radius $\frac{2}{b-1}$ centered at $-\frac{1}{b-1}$, while the region of convergence for the Laurent expansion is a circle of radius $\frac{1}{b-1}$ centered at 0. The minimum distance between the boundaries is again achieved at $Re(U)=0$, defining the region of convergence $(\frac{1}{b-1}i, \frac{\sqrt{3}}{b-1}i)$. Therefore, the minimax optimal threshold corresponds to $\alpha = \frac{1+\sqrt{3}}{2(b-1)}$.

\subsection{Degenerate case: $\beta = \gamma$}

In the degenerate case where the export rate is equal to the degradation rate, $U(s) = e^{-\gamma s}(u+\gamma vs)$. \\$|U|^2 = e^{-2\gamma s} (u+\gamma vs) (\bar{u} + \gamma \bar{v} s) \\ =  e^{-2\gamma s} (u\bar{u} +\gamma^2   v\bar{v} s^2 + \gamma [u\bar{v} + \bar{u}v] s) \\=  e^{-2\gamma s} (\gamma^2   |v|^2 s^2 + \gamma  [u\bar{v} + \bar{u}v] s +|u|^2 )$.

Note that $|U(s)| = 0$ only if $U(s) = 0$. This may occur if $u=v=0$ or $u=s=0$ which are trivial edge cases. The alternative case $u=-\gamma vs$ is not relevant since $\gamma, s>0$ and both $u$ and $v$ are only evaluated in the negative real half-plane. Thus, all nontrivial combinations of $u,v,s$ yield a strictly positive $|U(s)|$ whose behavior is one-to-one with that of $|U(s)|^2$.

Differentiation yields $\frac{d|U|^2}{ds} \\= 
e^{-2\gamma s}[2 \gamma^2 |v|^2 s + \gamma  [u\bar{v} + \bar{u}v]] - 2 \gamma e^{-2\gamma s}[\gamma^2 |v|^2 s^2 + \gamma  [u\bar{v} + \bar{u}v] s + |u|^2] 
\\= e^{-2\gamma s} [(-2\gamma^3 |v|^2) s^2 + (2 \gamma^2 |v|^2 -2\gamma^2  [u\bar{v} + \bar{u}v]) s + (\gamma  [u\bar{v} + \bar{u}v] -2\gamma|u|^2)] 
\\= 2\gamma e^{-2\gamma s} [(-\gamma^2 |v|^2) s^2 +( \gamma |v|^2-\gamma  [u\bar{v} + \bar{u}v])s + (\frac{1}{2} [u\bar{v} + \bar{u}v]-|u|^2 )]
\\= 2\gamma e^{-2\gamma s} [As^2 + Bs +C]$,

where $A=-\gamma^2|v|^2\le 0$, $B=\gamma ( |v|^2 - [u\bar{v} + \bar{u}v])$, and $C= \frac{1}{2} [u\bar{v} + \bar{u}v]-|u|^2 $

For $u=0$ or $v=0$, explicit solutions to $|U| = \alpha$ can be directly computed using the Lambert W function. For $u,v \ne 0$, a generalized Lambert W function is required for an analytical solution \cite{scott_general_2006, maignan_fleshing_2016}. Although approximation methods are available \cite{scott_asymptotic_2014, castle_taylor_2018}, they are not generally necessary here. Instead, a numerical scheme can be used to compute coarse estimates for roots constrained by the zeroes of $\frac{d|U|^2}{ds}$. As long as the roots are within $\Delta \alpha$ of $\alpha$, such that $(\alpha-\Delta \alpha, \alpha + \Delta \alpha)$ is within the radius of convergence for both expansions, the approximation is valid. 

\subsection{General case: $\gamma_e \ne \gamma_c$}

An analogous demonstration may be provided for the general case where the export rate is not equal to the degradation rate. 

As elsewhere, $f\equiv \frac{\beta}{\beta-\gamma}$. $U(s) = vf e^{-\gamma s} + (u-vf) e^{-\beta s}$.

$|U|^2 = [vf e^{-\gamma s} + (u-vf) e^{-\beta s}][\bar{v}f e^{-\gamma s} + (\bar{u}-\bar{v}f) e^{-\beta s}] \\= 
|vf|^2 e^{-2\gamma s} + |u-vf|^2 e^{-2\beta s} + [vf(\bar{u}-\bar{v}f) + \bar{v}f(u-vf)] e^{-(\gamma + \beta)s} \\=
|vf|^2 e^{-2\gamma s} + |u-vf|^2 e^{-2\beta s} + f[v(\bar{u}-\bar{v}f) + \bar{v}(u-vf)] e^{-(\gamma + \beta)s}  \\=
|vf|^2 e^{-2\gamma s} + |u-vf|^2 e^{-2\beta s} + f[v\bar{u}-|v|^2f + \bar{v}u-|v|^2f] e^{-(\gamma + \beta)s} \\=
f^2|v|^2 e^{-2\gamma s} + |u-vf|^2 e^{-2\beta s} + f([\bar{u}v + u\bar{v}]-2|v|^2f) e^{-(\gamma + \beta)s}$.

$|u-vf|^2 = (u-vf)(\bar{u} - \bar{v}f) = (|u|^2 + f^2|v|^2 - f[\bar{u}v + u\bar{v}])$.

Therefore, $|U|^2 = f^2|v|^2 e^{-2\gamma s} + (|u|^2 + f^2|v|^2 - f[\bar{u}v + u\bar{v}]) e^{-2\beta s} + f([\bar{u}v + u\bar{v}]-2|v|^2f) e^{-(\gamma + \beta)s}$.

Differentiation yields $\frac{d|U|^2}{ds} \\= 
-2 \gamma f^2|v|^2 e^{-2\gamma s} -2 \beta (|u|^2 + f^2|v|^2 - f[\bar{u}v + u\bar{v}]) e^{-2\beta s} - (\gamma+\beta) f([\bar{u}v + u\bar{v}]-2|v|^2f) e^{-(\gamma + \beta)s}
\\=0$ whenever
$C e^{-2\gamma s} + A e^{-2\beta s} + B e^{-(\gamma + \beta)s} =0$,

Where $C = -2 \gamma f^2|v|^2 < 0$, $A =-2 \beta (|u|^2 + f^2|v|^2 - f[\bar{u}v + u\bar{v}])<0$, and \\$B=-(\gamma+\beta) f([\bar{u}v + u\bar{v}]-2|v|^2f) \in {\rm I\!R}$.
Multiplying both sides by $e^{2\gamma s} (> 0)$,

$0=C + A e^{-2(\beta-\gamma )s} + B e^{-(\beta-\gamma )s} \\=
A e^{2(\gamma-\beta)s} + B e^{(\gamma-\beta) s} +C$

Making the substitution $z=e^{(\gamma-\beta)s}$,

$0= Az^2 + Bz + C$

Note that two roots exist, and the solutions $z^*$ may be converted back to the time domain per $s^* = \frac{1}{\gamma-\beta} \ln z^*$. Only real positive extrema $s^*$ are pertinent to the domain partition; all others may be discarded.

\subsection{Root computation}

Apart from the degenerate cases described above, where the roots can be computed \textit{via} logarithms or the Lambert $W$ function, the determination of domain boundaries requires the use of numerical routines. Specifically, we perform 20 iterations of the Newton-Raphson method to find locations where $|U|^2 = \alpha^2$. If the routine fails to converge, we switch to the bisection method and search in the domain bracketed by extrema of $|U|^2$; as discussed above, the locations of these extrema are known analytically. If the root is not bracketed (i.e., the rightmost extremum at the location $s^*$ has a value $y^*>\alpha^2$), we use the heuristic bracket $(s^*, s^* + 10/\min(\beta,\gamma))$, which suffices due to the fast (exponential) decay of the function $|U|^2$.

\section{Addenda}

\subsection{Solutions for arbitrary initial conditions}

As described by Singh and Bokes \cite{singh_consequences_2012}, the generating function at time $t$ for an arbitrary initial distribution is given by $\phi(u,v,t) = k_i \int_0^t (M(U(s))-1) ds + \phi(U(t),V(t),0)$, where $\phi(u,v,0)$ is the factorial cumulant generating function of the initial distribution. If $n(t=0) = n_0$ and $m(t=0) = m_0$, the corresponding generating function is $G(x,y,t=0) = \sum_{n} \sum_{m} x^{n} y^{m} P(n,m,t=0) = x^{n_m} y^{m_0}$. Its logarithm $\phi(u,v,0)$ is simply $n_0 \ln (1+u) + m_0 \ln (1+v)$. To find the value of this contribution, all that remains is to compute $U(t)$ and $V(t)$, where $U(s)$ is given above and $V(s) = ve^{-\gamma s}$. This method can be trivially extended to an arbitrary initial distribution of molecules by summing over the appropriate $n$ and $m$ and using the observed initial copy-number distribution as the initial condition $P(n,m,t=0)$.

\subsection{Negative binomial burst generating function}

A negative binomial (NB) burst distribution has $P(B=\rho) = \binom{\rho + a-1}{\rho}(1-p)^a p^\rho$, where $p = \frac{b}{1+b}$ and the mean burst size is $ba$. The resulting generating function is $F(x) = (\frac{1-p}{1-px})^a$. Substituting $M(u) = F(1+u)$ yields $M(u)-1 = \frac{1}{(1-bu)^a}-1$, which reduces to the geometric distribution for $a=1$. Defining $X=(bu)^{-1}$, the first term can be rewritten as $\frac{X^a}{X^a(1-X^{-1})^a} =\frac{X^a}{(X-1)^a}$. This form affords the Taylor expansion in the form of $\sum_{i=0}^\infty X^{i+a} (-1)^{i-a} \binom{-a}{i}$, and an analogous Laurent expansion. Therefore, it is, in principle, possible to compute the CME solutions through a series of integrals of the form $\int U^{i+a}ds$, $i\in \ZZ$, which are given by $\Gamma(i+a,\cdot)$ for the degenerate case and ${}_2 F_1 (i+a,\cdot;\cdot;\cdot)$ otherwise.

Previous discussions of the NB model treat it as a natural generalization of the geometric model, but do not motivate its use from physical arguments. Therefore, we do not develop the solution to the model in further detail.

\addcontentsline{toc}{section}{References}

\bibliography{references}